\documentclass[12pt]{article}
\usepackage{epsfig}
\def\be{\begin{equation}}
\def\ee{\end{equation}}
\def\bea{\begin{eqnarray}}
\def\eea{\end{eqnarray}}
\usepackage{graphicx}

\catcode`\@=11
\def\lsim{\mathrel{\mathpalette\@versim<}}
\def\gsim{\mathrel{\mathpalette\@versim>}}
\def\@versim#1#2{\vcenter{\offinterlineskip
\ialign{$\m@th#1\hfil##\hfil$\crcr#2\crcr\sim\crcr } }}
\catcode`\@=12

\catcode`@=12
\begin{document}
\thispagestyle{empty}
\begin{flushright}
UCRHEP-T581\\
July 2017\
\end{flushright}
\vspace{0.6in}
\begin{center}
{\LARGE \bf Cobimaximal Neutrino Mixing from $S_3 \times Z_2$\\}
\vspace{1.2in}
{\bf Ernest Ma$^{a,b}$\\}
\vspace{0.2in}
{\sl $^a$ Physics \& Astronomy Department and Graduate Division,\\ 
University of California, Riverside, California 92521, USA\\}
\vspace{0.1in}
{\sl $^b$ HKUST Jockey Club Institute for Advanced Study,\\
Hong Kong University of Science and Technology, Hong Kong, China\\}
\end{center}
\vspace{1.2in}

\begin{abstract}\
It has recently been shown that the phenomenologically successful pattern of 
cobimaximal neutrino mixing ($\theta_{13} \neq 0$, $\theta_{23} = \pi/4$, 
and $\delta_{CP} = \pm \pi/2$) may be achieved in the context of 
the non-Abelian discrete symmetry $A_4$.  In this paper, the same goal 
is achieved with $S_3 \times Z_2$.  The residual 
lepton $Z_3$ triality in the case of $A_4$ is replaced here by 
$Z_2 \times Z_2$.  The associated phenomenology of the scalar sector is 
discussed. 
\end{abstract}

\newpage
\baselineskip 24pt
\noindent \underline{Introduction}:

Present neutrino data~\cite{pdg2014,t2k15} are indicative of 
$\theta_{13} \neq 0$, 
$\theta_{23} = \pi/4$ and $\delta_{CP} = -\pi/2$.  Calling this 
\underline{cobimaximal} mixing~\cite{m15}, it has been shown that it may 
be derived in two equivalent ways. (I) The Majorana neutrino mass matrix, 
in the basis where charged-lepton masses are diagonal, is of the 
form~\cite{m02,bmv03,gl04}
\begin{equation}
{\cal M}_\nu^{(e,\mu,\tau)} = \pmatrix{A & C & C^* \cr C & D^* & B \cr 
C^* & B & D},
\end{equation}
where $A,B$ are real. (II) The neutrino mixing matrix is of the 
form~\cite{fmty00,mty01,m15-1,h15}
\begin{equation}
U_{l \nu} = U_\omega {\cal O},
\end{equation}
where~\cite{c78,w78}
\begin{equation}
U_\omega = {1 \over \sqrt{3}} \pmatrix{1 & 1 & 1 \cr 1 & \omega & \omega^2 \cr 
1 & \omega^2 & \omega},
\end{equation}
with $\omega = \exp(2 \pi i/3) = -1/2 + i \sqrt{3}/2$, and ${\cal O}$ is 
any arbitrary real orthogonal matrix.  This yields $|U_{\mu i}| = 
|U_{\tau i}|$ which leads to cobimaximal mixing.  Using the fact that 
$U_\omega$ is 
derivable from $A_4$~\cite{mr01}, and the scotogenic generation of 
neutrino mass from a set of real scalars~\cite{m15-1,fmp14,fmz16,m16}, Eq.~(2) 
is naturally achieved.  This conceptual shift from 
tribimaximal~\cite{hps02,m04} to cobimaximal mixing may also be understood 
as the result of a residual generalized $CP$ 
symmetry~\cite{gl04,cyd15,mn15,jp15,hrx15,n16,lld16}.  Here we 
show how cobimaximal mixing may be obtained from the soft breaking of 
$S_3 \times Z_2$ to $Z_2 \times Z_2$ instead of the soft breaking of 
$A_4$ to $Z_3$.

\noindent \underline{Soft breaking of $S_3$ to $Z_2$ with two Higgs doublets}: 

Let $(\Phi_1,\Phi_2) \sim \underline{2}$ under $S_3$, then the most general 
$S_3$ invariant scalar potential for $\Phi_{1,2}$ is given by~\cite{cfm04}
\begin{eqnarray}
V_0 &=& \mu_0^2 (\Phi_1^\dagger \Phi_1 + \Phi_2^\dagger \Phi_2) 
+ {1 \over 2} \lambda_1 (\Phi_1^\dagger \Phi_1 + \Phi_2^\dagger \Phi_2)^2 
\nonumber \\ 
&+& {1 \over 2} \lambda_2 (\Phi_1^\dagger \Phi_1 - \Phi_2^\dagger \Phi_2)^2 
+ \lambda_3 (\Phi_1^\dagger \Phi_2)(\Phi_2^\dagger \Phi_1).
\end{eqnarray}
The soft breaking of $V_0$ to a residual $Z_2$ symmetry may be accomplished 
in three ways, according to how the $Z_2$ is chosen.  (1) $\Phi_1 \to 
\Phi_1$, $\Phi_2 \to -\Phi_2$.  The soft term required is then 
$\Phi_1^\dagger \Phi_1 - \Phi_2^\dagger \Phi_2$.  (2) $\Phi_1 \to \Phi_2$, 
$\Phi_2 \to \Phi_1$.  The soft term required is then $\Phi_1^\dagger \Phi_2 
+ \Phi_2^\dagger \Phi_1$. (3) $\Phi_1 \to i \Phi_2$, $\Phi_2 \to -i \Phi_1$. 
The soft term required is then $i (\Phi_1^\dagger \Phi_2 - \Phi_2^\dagger 
\Phi_1)$.  In (1), the residual $Z_2$ symmetry corresponds to $\langle 
\phi_1^0 \rangle \neq 0$,  $\langle \phi_2^0 \rangle = 0$.
In (2), it is $\langle \phi_1^0 + \phi_2^0 \rangle \neq 0$,  
$\langle \phi_1^0 - \phi_2^0 \rangle = 0$.
In (3), it is $\langle \phi_1^0 + i\phi_2^0 \rangle \neq 0$,  
$\langle \phi_1^0 - i\phi_2^0 \rangle = 0$.

\noindent \underline{Two lepton families under $S_3$}:

Let $(\nu_1,l_1)_L, (\nu_2,l_2)_L \sim \underline{2}$ under $S_3$, and 
$l_{1R},l_{2R} \sim \underline{1}', \underline{1}$ under $S_3$.  The 
$S_3$ invariant Yukawa terms for charged-lepton masses are then
\begin{equation}
-{\cal L}_Y = f_\mu (\bar{l}_{1L} \phi_1^0 - \bar{l}_{2L} \phi_2^0) l_{1R} 
+ f_\tau (\bar{l}_{1L} \phi_1^0 + \bar{l}_{2L} \phi_2^0) l_{2R} + H.c.
\end{equation}
Choosing the third option for $Z_2$ with $i \mu_{12}^2 (\Phi_1^\dagger \Phi_2 
- \Phi_2^\dagger \Phi_1)$ with $\mu_{12}^2 < 0$, so that $v/\sqrt{2} = \langle 
\phi_1^0 \rangle = i \langle \phi_2^0 \rangle$, the $2 \times 2$ 
mass matrix linking $(\bar{l}_{1L}, \bar{l}_{2L})$ to $(l_{1R},l_{2R})$ 
is then given by
\begin{equation}
{\cal M}_l = \pmatrix{f_\mu & f_\tau \cr i f_\mu & -i f_\tau} {v \over 
\sqrt{2}} = {1 \over \sqrt{2}} 
\pmatrix{1 & 1 \cr i & -i} \pmatrix{f_\mu v & 0 \cr 0 & f_\tau v}.
\end{equation}

\noindent \underline{Three lepton families under $S_3 \times Z_2$}:

A third lepton family may be added which transforms as $(\underline{1},-)$ 
under $S_3 \times Z_2$, so that it couples to a third Higgs doublet 
which transforms as $(\underline{1},+)$.  The $3 \times 3$ unitary matrix 
linking the diagonal charged-lepton mass matrix to the neutrino mass 
matrix is then
\begin{equation}
U_2 = \pmatrix{1 & 0 & 0 \cr 0 & 1/\sqrt{2} & -i/\sqrt{2} \cr 0 & 1/\sqrt{2} 
& i/\sqrt{2}}.
\end{equation}
This serves the same purpose as $U_\omega$ of Eq.~(3), because 
\begin{equation}
U_{l \nu} = U_2 {\cal O}
\end{equation}
also yields $|U_{\mu i}| = |U_{\tau i}|$ which leads to cobimaximal mixing.
In fact, since $U_{ei} = {\cal O}_{1i}$, the $\theta_{12}$ and $\theta_{13}$ 
angles are the same in both $U_{l\nu}$ and ${\cal O}$.

\noindent \underline{More about ${\cal O}$}:

The $3 \times 3$ Majorana neutrino mass matrix ${\cal M}_\nu$ is symmetric 
but also complex in general with three physical phases.  It is thus not 
diagonalized by an orthogonal matrix.  However, if the origin of this mass 
matrix is radiative and comes from a set of three real scalars, and there 
are no extraneous phases from the additional interactions, then it is 
possible~\cite{m15-1,fmp14,fmz16,m16} to achieve this result. 

In gneral ${\cal M}_\nu$ is diagonalized by a unitary matrix with 3 angles 
and 3 phases:
\begin{equation}
U = \pmatrix{1 & 0 & 0 \cr 0 & c_{23} & s_{23} \cr 0 & -s_{23} & c_{23}} 
\pmatrix{c_{13} & 0 & s_{13}e^{-i \delta} \cr 0 & 1 & 0 \cr -s_{13} 
e^{i \delta} & 0 & c_{13}} \pmatrix{c_{12} & s_{12} & 0 \cr -s_{12} & c_{12} 
& 0 \cr 0 & 0 & 1} \pmatrix{1 & 0 & 0 \cr 0 & e^{i \alpha_{21}/2} & 0 \cr 
0 & 0 & e^{i \alpha_{31}/2}},
\end{equation}
where $c_{ij} = \cos \theta_{ij}$ and $s_{ij} = \sin \theta_{ij}$. 
If $\delta = 0$, then it is equal to an orthogonal matrix times a diagonal 
matrix involving only Majorana phases.  Upon multiplification on the left by 
$U_2$ of Eq.~(7), it will still lead to cobimaximal neutrino mixing. 
Now
\begin{equation}
U_2 \pmatrix{1 & 0 & 0 \cr 0 & c_{23} & s_{23} \cr 0 & -s_{23} & c_{23}} 
= \pmatrix{1 & 0 & 0 \cr 0 & e^{i \theta_{23}} & 0 \cr 0 & 0 & 
-e^{-i \theta_{23}}} \pmatrix{1 & 0 & 0 \cr 0 & 1/\sqrt{2} & -i/\sqrt{2} \cr 
0 & -1/\sqrt{2} & -i/\sqrt{2}}.
\end{equation}
The diagonal matrix of phases on the left may be absorbed into the charged 
leptons, and the remaining part of $U_2 U$ becomes
\begin{equation}
\pmatrix{1 & 0 & 0 \cr 0 & 1/\sqrt{2} & 1/\sqrt{2} \cr 0 & -1/\sqrt{2} & 
1/\sqrt{2}} 
\pmatrix{c_{13} & 0 & is_{13}e^{-i \delta} \cr 0 & 1 & 0 \cr is_{13} 
e^{i \delta} & 0 & c_{13}} \pmatrix{c_{12} & s_{12} & 0 \cr -s_{12} & c_{12} 
& 0 \cr 0 & 0 & 1} \pmatrix{1 & 0 & 0 \cr 0 & e^{i \alpha_{21}/2} & 0 \cr 
0 & 0 & -ie^{i \alpha_{31}/2}}.
\end{equation}
This means that if $\delta=0$, cobimaximal mixing is achieved with 
$e^{-i \delta_{CP}} = e^{i\pi/2} = i$ as expected.  However, even if 
$\delta \neq 0$, so that $\delta_{CP}$ deviates from $-\pi/2$,  
$\theta_{23}$ remains at $\pi/4$.  This is a remarkable result and it is only 
true because of $U_2$ of Eq.~(7), and does not hold for $U_\omega$ of Eq.~(3).
The deviation from cobimaximal mixing is model-dependent and has been 
calculated~\cite{mnp15} in a specific model, showing the correlation of 
$\delta_{CP}$ with $\theta_{23}$.  Here only $\delta_{CP}$ deviates.

\noindent \underline{Soft breaking of $S_3$ to $Z_2$ with three Higgs 
doublets}: 

Adding $\Phi_3 \sim \underline{1}$ under $S_3$, the scalar potential of 
our model becomes
\begin{eqnarray}
V &=& \mu_0^2 (\Phi_1^\dagger \Phi_1 + \Phi_2^\dagger \Phi_2) 
+ i \mu^2_{12} (\Phi_1^\dagger \Phi_2 - \Phi_2^\dagger \Phi_1) 
+ \mu_3^2 \Phi_3^\dagger \Phi_3 + \left[ {1 \over 2} \mu^2_{30} \Phi_3^\dagger 
(\Phi_1 + i \Phi_2) + H.c. \right] \nonumber \\ 
&+& {1 \over 2} \lambda_1 (\Phi_1^\dagger \Phi_1 + \Phi_2^\dagger \Phi_2)^2 
+ {1 \over 2} \lambda_2 (\Phi_1^\dagger \Phi_1 - \Phi_2^\dagger \Phi_2)^2 
+ \lambda_3 (\Phi_1^\dagger \Phi_2)(\Phi_2^\dagger \Phi_1) + {1 \over 2} 
\lambda_4 (\Phi_3^\dagger \Phi_3)^2 \nonumber \\ 
&+& \lambda_5 (\Phi_3^\dagger \Phi_3)(\Phi_1^\dagger \Phi_1 + \Phi_2^\dagger 
\Phi_2) + \lambda_6 \Phi_3^\dagger (\Phi_1 \Phi_1^\dagger + \Phi_2 
\Phi_2^\dagger) \Phi_3 + 
[\lambda_7 (\Phi_3^\dagger \Phi_1)(\Phi_3^\dagger \Phi_2) + H.c.]
\end{eqnarray}
The $\mu^2_{12}$ and $\mu_{30}^2$ terms break $S_3$ softly to $Z_2$, under 
which $\Phi_3$ and $\Phi_+ = (\Phi_1 + i \Phi_2)/\sqrt{2}$ are even and 
$\Phi_- = (\Phi_1 - i \Phi_2)/\sqrt{2}$ is odd.  The $S_3$ allowed quartic 
term $(\Phi_3^\dagger \Phi_1)(\Phi_2^\dagger \Phi_1) + 
(\Phi_3^\dagger \Phi_2)(\Phi_1^\dagger \Phi_2)$ is forbidden by 
making $\Phi_3$ odd under an extra $Z_2$ symmetry, which is then broken 
softly by the $\mu^2_{30}$ term.  With this modification, $(\nu_e,e)_L$ is 
even and $e_R$ is odd under this softly broken extra $Z_2$ to allow the 
Yukawa coupling $f_e \bar{e}_L e_R \phi_3^0$, with $m_e = f_e v_3$.    
Assuming a small $\mu^2_{30}$ term, $v_3$ is naturally much smaller than 
$v$.  Hence $m_e << m_\mu, m_\tau$ and 
the charged leptons are distinguished from each other according to
\begin{equation}
e \sim (+,-), ~~~ \mu \sim (-,+), ~~~ \tau \sim (+,+).
\end{equation}
The $Z_3$ triality~\cite{m10,cdmw11} coming from $A_4$ has now been 
replaced by the above under $Z_2 \times Z_2$.  This serves to forbid 
$\mu \to e \gamma$, etc. as in the case of $Z_3$ lepton triality. 
The odd Higgs doublet $\Phi_-$ transforms as $(-,+)$ and couples 
to $\bar{\mu}_L \tau_R$ and $\bar{\tau}_L \mu_R$ as in Ref.~\cite{mm13}.

\noindent \underline{Phenomenology of scalar interactions}:

The leptonic Yukawa interactions are given by
\begin{eqnarray}
-{\cal L}_Y &=& f_\tau (\bar{\tau}_L \phi_+^0 + \bar{\mu}_L \phi_-^0) \tau_R 
+ f_\tau (\bar{\nu}_\tau \phi_+^+ + \bar{\nu}_\mu \phi_-^+) \tau_R \nonumber \\
&+& f_\mu (\bar{\mu}_L \phi_+^0 + \bar{\tau}_L \phi_-^0) \mu_R 
+ f_\mu (\bar{\nu}_\mu \phi_+^+ + \bar{\nu}_\tau \phi_-^+) \mu_R \nonumber \\
&+& f_e \bar{e}_L \phi_3^0 e_R + f_e \bar{\nu}_e \phi_3^+ e_R + H.c.
\end{eqnarray}
The scalar interactions are given by
\begin{eqnarray}
V &=& (\mu^2_0 + \mu^2_{12}) \Phi_+^\dagger \Phi_+ + (\mu^2_0 - \mu^2_{12}) 
\Phi_-^\dagger \Phi_- + \mu^2_3 \Phi_3^\dagger \Phi_3 + \left[ {1 \over 
\sqrt{2}} \mu^2_{30} \Phi_3^\dagger \Phi_+ + H.c. \right] \nonumber \\ 
&+& \left( {1 \over 2} \lambda_1 + {1 \over 4} \lambda_3 \right) 
(\Phi_+^\dagger \Phi_+)^2 + \left( {1 \over 2} \lambda_1 + {1 \over 4} 
\lambda_3 \right) (\Phi_-^\dagger \Phi_-)^2 + \left( \lambda_1 - {1 \over 2} 
\lambda_3 \right) (\Phi_+^\dagger \Phi_+)(\Phi_-^\dagger \Phi_-) \nonumber \\ 
&+& \left( {1 \over 2} \lambda_2 - {1 \over 4} \lambda_3 \right) 
[(\Phi_+^\dagger \Phi_-)^2 + (\Phi_-^\dagger \Phi_+)^2] + \left( \lambda_2 
+ {1 \over 2} \lambda_3 \right) ( \Phi_+^\dagger \Phi_-)(\Phi_-^\dagger 
\Phi_+) \nonumber \\ &+& {1 \over 2} \lambda_4 (\Phi_3^\dagger \Phi_3)^2 
+ \lambda_5 (\Phi_3^\dagger \Phi_3) (\Phi_+^\dagger \Phi_+ + \Phi_-^\dagger 
\Phi_-) + \lambda_6 \Phi_3^\dagger ( \Phi_+ \Phi_+^\dagger + \Phi_- 
\Phi_-^\dagger) \Phi_3 \nonumber \\ 
&+& \left( {1 \over 2i} \lambda_7 [(\Phi_3^\dagger \Phi_+)^2
- (\Phi_3^\dagger \Phi_-)^2] + H.c. \right)
\end{eqnarray}
Assuming $v,v_3$ to be real, the conditions for minimizing $V$ are
\begin{eqnarray}
&& v [(\mu_0^2 + \mu_{12}^2) + (\lambda_1 + {1 \over 2} \lambda_3) v^2 + 
(\lambda_5 + \lambda_6 + Im (\lambda_7)) v_3^2] + {1 \over \sqrt{2}} \mu_{30}^2 
v_3 = 0, \\
&& v_3 [\mu_3^2 + \lambda_4 v_3^2 + (\lambda_5 + \lambda_6 + Im (\lambda_7)) 
v^2] + {1 \over \sqrt{2}} \mu_{30}^2 v = 0.
\end{eqnarray}
For $v_3 << v$, we obtain
\begin{eqnarray}
v^2 &\simeq& {-(\mu_0^2 + \mu_{12}^2) \over \lambda_1 + (\lambda_3/2)}, \\ 
v_3 &\simeq& {-\mu_{30}^2 v \over \sqrt{2} [\mu_3^2 + (\lambda_5 + \lambda_6 
+ Im (\lambda_7))v^2]}.
\end{eqnarray}
The states $\sqrt{2}[v Im(\phi_+^0) + v_3 Im(\phi_3^0)]/\sqrt{v^2+v^3}$ and 
$[v \phi_+^\pm + v_3 \phi_3^\pm]/\sqrt{v^2 + v_3^2}$ are 
the would-be massless Goldstone modes for the $Z$ and $W^\pm$ bosons. 
The states $A = \sqrt{2}[v Im(\phi_3^0) - v_3 Im(\phi_+^0)]/\sqrt{v^2+v_3^2}$ 
and $H^\pm = [v \phi_3^\pm - v_3 \phi_+^\pm]/\sqrt{v^2+v_3^2}$ 
have masses given by
\begin{eqnarray}
m_A^2 &=& - Im(\lambda_7) (v^2 + v_3^2) - { \mu_{30}^2 (v^2 + v_3^2) \over 
\sqrt{2} v v_3} \simeq \mu_3^2 + (\lambda_5 + \lambda_6 - Im(\lambda_7))v^2, \\ 
m_{H^\pm}^2 &=& - (\lambda_6 + Im(\lambda_7)) (v^2 + v_3^2) - {\mu_{30}^2 
(v^2 + v_3^2) \over \sqrt{2} v v_3} \simeq \mu_3^2 + \lambda_5 v^2.
\end{eqnarray}
The states $h = \sqrt{2} Re(\phi_+^0)$ and $H = \sqrt{2}Re(\phi_3^0)$ are 
approximate mass eigenstates with 
\begin{equation}
m_h^2 \simeq (2 \lambda_1 + \lambda_3) v^2, ~~~ m_H^2 \simeq \mu_3^2 + 
(\lambda_5 + \lambda_6 + Im(\lambda_7)) v^2,
\end{equation}
and $h-H$ mixing given by
\begin{equation}
\epsilon \simeq {- v_3 \over v} \left[ {\mu_3^2 - (\lambda_5 + \lambda_6 
+ Im(\lambda_7)) v^2 \over \mu_3^2 + (\lambda_5 + \lambda_6 
+ Im(\lambda_7)) v^2} \right].
\end{equation}
The $\Phi_-$ doublet has odd $Z_2$ and does not mix with $\Phi_+$ or $\Phi_3$. 
The masses of its components are given by
\begin{eqnarray}
m^2(\phi_-^\pm) &\simeq& \mu_0^2 - \mu_{12}^2 + \left( \lambda_1 - {1 \over 2} 
\lambda_3 \right) v^2, \\
m^2(\sqrt{2} Re(\phi_-^0)) &\simeq& \mu_0^2 - \mu_{12}^2 + \left( \lambda_1 
- {1 \over 2} \lambda_3 + 2 \lambda_2 \right) v^2, \\
m^2(\sqrt{2} Im(\phi_-^0)) &\simeq& \mu_0^2 - \mu_{12}^2 + \left( \lambda_1 
+ {1 \over 2} \lambda_3 \right) v^2.
\end{eqnarray}

\noindent \underline{Phenomenology of lepton interactions}:

From Eq.~(14), the lepton interactions of this model are given by
\begin{eqnarray}
-{\cal L}_Y &=& {m_\tau \over v \sqrt{2}} h \bar{\tau} \tau + {m_\mu \over 
v \sqrt{2}} h \bar{\mu} \mu + {m_e \over v_3 \sqrt{2}} [(H + iA) \bar{e}_L e_R 
+ H^+ \bar{\nu_e} e_R + H.c.] \nonumber \\ 
&+& \left[ {m_\tau \over v} [\phi_-^0 \bar{\mu}_L \tau_R + \phi_-^+ \bar{\nu_\mu} 
\tau_R] + {m_\mu \over v} [\phi_-^0 \bar{\tau}_L \mu_R + \phi_-^+ \bar{\nu_\tau} 
\mu_R] \right] + H.c. 
\end{eqnarray}
to a very good approximation.  Since $v_3 << v$ is assumed, the heavy $H$ 
and $A$ couple predominantly to $e^- e^+$.  If they are produced, through 
a virtual $Z$ for example, at the Large Hadron Collider (LHC), the 
$e^-e^+e^-e^+$ final state is very distinctive and potentially measurable. 
In the same way, $\sqrt{2}Re(\phi_-^0) + \sqrt{2}Im(\phi_-^0)$ may be produced. 
They decay to $\mu^- \tau^+$ and $\mu^+ \tau^-$ which are again rather 
distinctive if $\tau^\pm$ can be reconstructed experimentally. 
On the other hand, the decay of $\phi_-^\pm$ is predominantly to 
$\tau^+ \nu_\mu$, $\tau^- \bar{\nu}_\mu$.

\noindent \underline{Conclusion}:

The notion of cobimaximal neutrino mixing, i.e. $\theta_{13} \neq 0$, 
$\theta_{23} = \pi/4$, and $\delta_{CP} = -\pi/2$, is shown to be a 
consequence of the residual $Z_2 \times Z_2$ symmetry of an $S_3 \times Z_2$ 
model of lepton masses.  This is an alternative theoretical understanding 
from the usual $A_4$ realization.  It has verifiable decay signatures 
in its three Higgs doublets, as well as the prediction that even if 
$\delta_{CP}$ deviates from $-\pi/2$, $\theta_{23}$ will remain at $\pi/4$, 
in contrast to other models.

\noindent \underline{Acknowledgment}:
This work is supported in part 
by the U.~S.~Department of Energy under Grant No.~DE-SC0008541.

\bibliographystyle{unsrt}

\end{document}